\documentstyle[epsfig]{mn}

\newif\ifAMStwofonts
\AMStwofontstrue

\def\be{\begin{equation}}
\def\ee{\end{equation}}

\def\gtsima{$\; \buildrel > \over \sim \;$}
\def\ltsima{$\; \buildrel < \over \sim \;$}
\def\prosima{$\; \buildrel \propto \over \sim \;$}
\def\gsim{\lower.5ex\hbox{\gtsima}}
\def\lsim{\lower.5ex\hbox{\ltsima}}
\def\simgt{\lower.5ex\hbox{\gtsima}}
\def\simlt{\lower.5ex\hbox{\ltsima}}
\def\simpr{\lower.5ex\hbox{\prosima}}

\def\etal{{\it et al.~}}

\title{How is the Reionization Epoch Defined ?}

\author[M. Bruscoli, A. Ferrara \& E. Scannapieco]{Marialuce Bruscoli$^1$, 
Andrea Ferrara$^2$ \& Evan Scannapieco$^2$\\
$^1$Dipartimento di Astronomia, Universit\`a degli studi 
di Firenze, L.go E. Fermi 2, Firenze, Italy\\
$^2$Osservatorio Astrofisico di Arcetri, L.go E. Fermi 5,
Firenze, Italy\\}
\pagerange{\pageref{firstpage}--\pageref{lastpage}}
\pubyear{2001}
\begin{document}

\maketitle
\label{firstpage}

\begin{abstract}

We study the effect of a prolonged epoch of reionization on the
angular power spectrum of the Cosmic Microwave Background.  Typically
reionization studies assume a sudden phase transition, with the
intergalactic gas moving from a fully neutral to a fully ionized state
at a fixed redshift. Such models are at odds, however, with detailed
investigations of reionization, which favor a more extended
transition.  We have modified the code {\tt CMBFAST} to allow the
treatment of more realistic reionization histories and applied it to
data obtained from numerical simulations of reionization.  We show
that the prompt reionization assumed by {\tt CMBFAST} in its original
form heavily contaminates any constraint derived on the reionization
redshift.  We find, however, that prompt reionization models give a
reasonable estimate of the epoch at which the mean cosmic ionization
fraction was $\approx 50 \%$, and provide a very good measure of the
overall Thomson optical depth.  The overall differences in the
temperature (polarization) angular power spectra between prompt and
extended models with equal optical depths are less than $1\%$
($10\%$).

\end{abstract}
 

\section{INTRODUCTION}
 
At a redshift of $z\approx 1100$ the intergalactic medium (IGM)
recombined and remained neutral until the first sources of ionizing
radiation formed.  While the distribution and evolution of these
sources are unknown, the overall process of IGM reionization is fairly
well understood, and can be divided into three phases.  First,
individual HII regions developed around the sources. Then, these
ionized regions grew in number and size until they overlapped,
producing a sudden increase of the photon mean free path.  Finally,
when all underdense regions and voids were completely ionized,
photons penetrated into overdense clumps and filaments, 
bringing the reionization process to completion at a ``reionization
redshift.''

It is clear that the mystery of the nature and evolution of the
ionizing sources is intimately tied with the time scale over
which these phases took place. Thus there is a great deal of physical
information that could be gathered if the reionization redshift,
$z_i$, and its duration, $\Delta z$, can be firmly established.

Apart from the study of the (HI and HeII) Gunn-Peterson effect, which
has not yet yielded conclusive results (Becker \etal 2001; Gnedin 2001; 
Theuns \etal 2001), measurements of Cosmic Microwave Background (CMB) 
anisotropies have the greatest potential for constraining these important 
quantities (Griffiths \etal 1999). 
The scattering of CMB photons by free electrons damps the angular power 
spectrum of primary anisotropies by a factor of $e^{-2 \tau}$ for large
angular multipoles $\ell \gsim 100$ (Tegmark \&
Zaldarriaga 2000), where $\tau$ is the Thomson optical depth.
Using the currently available data, consistency with the lack of 
Gunn-Peterson trough and with the observed
peak in the angular power spectrum at $\ell \approx 200$ (De Bernardis
\etal 2000; Hanany \etal 2000; Padin \etal 2001) is able to constrain
$0.02 \le \tau \le 0.44$ (De Bernardis \etal 1997; Griffiths \etal
1999; Tegmark \etal 2001; Griffiths \& Liddle 2001),
although these results are somewhat dependent on the cosmological
model assumed. Much better
constraints are expected from polarization studies to be carried out
by future satellites such as
 {\it
SPOrt}\footnote{http://sport.tesre.bo.cnr.it} and {\it
PLANCK}\footnote{http://astro.estec.esa.nl/SA-general/Projects/Planck}.

A second method of quantifying reionization is to examine $z_i$, and
several recent studies have attempted to use the CMB to derive 
this redshift directly, using codes that assume a sudden epoch of
reionization (e.g, Hu \etal 95; Seljak \& Zaldarriaga 1996).  For
example, Schmalzing \etal (2000) use {\it MAXIMA} data in combination
with cosmological parameters from independent measurements of Big Bang
Nucleosynthesis and X-ray cluster data to constrain $z_i$. By
performing a $\chi^2$ analysis, they conclude that $z_i > 15 (8)$ at
the 68\% (95\%) confidence level.  Similarly, Naselsky \etal (2001),
in addition to providing support to Schmalzing \etal results, showed
that polarization spectra are very sensitive to the reionization
redshift.

An important caveat is implicit in these applications however, namely
the assumption of prompt reionization.  While taking $\Delta z = 0$ is
a reasonable first step, more detailed simulations show that this
approach paints a picture of reionization in only the broadest of
strokes.  This is particularly true as the onset of reionization
raises the overall temperature of the IGM, suppressing the further
formation of objects that are too weakly bound gravitationally to
overcome the drastic increase in thermal pressure (Barkana \& Loeb
1999).  Thus recent numerical simulations of reionization (Ciardi
\etal 2000, hereafter CFGJ; Bruscoli \etal 2000; Gnedin 2000; Benson
\etal 2000) have shown that the evolution of the mean ionization
fraction is slow, has a nonlinear dependence on redshift, and occurs
in an extremely patchy manner. All of these details may considerably
affect the determination of $z_i$.

In this paper, we aim to clarify the effects of realistic reionization
scenarios on the observables typically used to examine the
reionization epoch.  To this end we have modified the most heavily
relied on theoretical code for computing CMB fluctuations, {\tt
CMBFAST} (Seljak \& Zaldarriaga 1996), to allow the treatment of such
reionization histories, helping to clarify the best approach in trying
to quantify and define the epoch of reionization.

The structure of this work is as follows.  In \S 2 we discuss
numerical models of reionization and how these have been
incorporated into the {\tt CMBFAST} code.  In \S 3 we apply
these techniques to study the impact of these models on the
temperature and polarization spectrum of the CMB, and conclusions
are given in \S 4.

\section{Reionization history from simulations}

To assess the effects of more physical reionization histories on the
determination of $z_i$, we have modified the Boltzmann code {\tt
CMBFAST} such that the redshift evolution of the volume-averaged mean
hydrogen ionization fraction, $x_{e}(z)$, can be specified, reading in data
from more detailed simulations of 
reionization.\footnote{A copy of this version of {\tt CMBFAST }is 
publicly available
at {\tt http://www.arcetri.astro.it/science/cosmology/CMB}}
Our modified version of {\tt CMBFAST} is then able to consider three
types of reionization histories: ({\it i}) prompt reionization models in 
which the
total optical depth, $\tau$, is chosen ({\it ii}) prompt models in
which the reionization redshift, $z_{i}$, is fixed, and ({\it iii})
models in which $x_{e}(z)$ is read from an external file.

As in Bruscoli \etal (2000), we consider a set of reionization
simulations studied in CFGJ, which model reionization from stellar
sources, including Population III objects.  These simulations were
developed in a critical density CDM universe with cosmological
parameters $h=0.5$, $\sigma_8=0.6$, and $\Omega_b=0.06$ (where $h$ is
the Hubble constant in units of $100 \, {\rm km} \, {\rm s}^{-1} \,
{\rm Mpc}^{-1}$, $\sigma_8$ is the rms variation of density
perturbations at the $8h^{-1}$ comoving Mpc length scale, and
$\Omega_b$ is the baryonic density in units of the critical density)
but span a large range of model parameters, and thus are suitable
for drawing conclusions as to the effect of extended reionization
in more general cosmologies.

The spatial distribution of the sources and ionized regions at
various cosmic epochs and the overall evolution of $x_{e}(z)$ are
obtained from high-resolution cosmological simulations within a
periodic box of comoving length $L=2.55h^{-1}$ Mpc. The source
properties are calculated taking into account a self-consistent
treatment of both radiative feedback from ionizing and
H$_2$--photodissociating photons and stellar feedback regulated by
supernovae from massive stars.  There are two main free parameters in
the simulations: the fraction of total baryons converted into stars,
$f_{b\star}$, and the escape fraction of ionizing photons from a given
galaxy, $f_{esc}$.  A critical discussion of these parameters is given
in CFGJ, and here we examine four different models, fixing
($f_{b\star}=0.012, f_{esc}=0.2$) in run A, (0.004, 0.2) in run B,
(0.15, 0.2) in run C , and (0.012, 0.1) in run D.  As run A, in which
$z_i \approx 11$, gives the best agreement between the derived
evolution of the cosmic star formation rate and observations at
$z\simlt 4$ (Steidel \etal 1998, CFGJ), we choose this as
our fiducial model, and rely on runs $B$, $C$, and $D$ to study the
effects of model uncertainties on our conclusions.

\section{Prompt vs. Simulated Reionization}

Equipped with this more general tool, we now quantitatively
contrast the prompt reionization $x_{e,p}(z)$ with the one derived
from the simulations, $x_{e,s}(z)$.  As {\tt CMBFAST} can be run either by 
assigning $z_i$ or $\tau$, we are able to examine both these cases separately.

\subsection{Assigning the Reionization Redshift}

\begin{figure}
\centerline{
\epsfig{figure=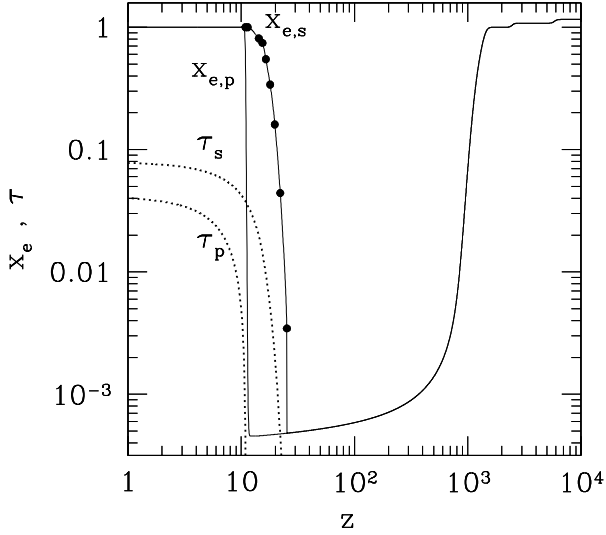, height=7cm}}
\caption{{\protect\footnotesize {H ionization fraction, $x_{e}$, and Thomson 
optical depth, $\tau$, versus redshift 
for run A (solid lines) and for prompt reionization (dotted 
lines) when {\tt CMBFAST } is run choosing $z_i=10.9$. The dots represent the 
output from the simulation.}}}
\label{fig1}
\end{figure}

In Fig. \ref{fig1} we plot the redshift evolution of the hydrogen
ionization fraction and the Thomson optical depth for the prompt and
simulated cases, choosing in the prompt case the same value of $z_i =
10.9$ as deduced from run A.  In this plot, the smoother, nonlinear
increase of $x_{e,s}(z)$ is evident, with $\Delta z \approx 15$.  Note
also that the optical depth for the two reionization histories are
quite different, and that $\tau_s=0.08$ for run A and $\tau_p=0.04$
for prompt reionization.
If we define the discrepancy between the prompt and simulated
visibility functions $g(z)$ as
\begin{equation}
\frac{\Delta g}{g}=\frac{|g_s(z)-g_p(z)|}{g_s(z)} ,
\label{eq1}
\end{equation}
where $g(z)$ is the probability that a photon reaching the observer
last scattered in the redshift range $z$ and $z+dz$, then we find that
$\Delta g/g \approx 5 \%$ for a large range of redshifts
between recombination and the beginning of reionization ($25 \simlt z
\simlt 1000$).

The most important comparison between the prompt and fiducial cases,
however, is made in terms of the different resulting CMB temperature
and polarization angular power spectra, $C_\ell$ and $P_\ell$, as
these are the observable quantities.  These are shown in the upper
left panel of Fig.\ 2, in which differences are particularly evident in
the polarization spectra, which are more sensitive to the reionization
history.  Reionization introduces a characteristic bump in the lower
multipoles ($\ell= 5 - 10$) that tends to shift to higher $\ell$'s
when $\tau$ is increased. The amplitude of the bump in the $P_\ell$
spectrum decreases with $\tau$ and therefore the signal is weaker for
the prompt reionization case. In particular the bump amplitude in the
angular spectrum $P_\ell$ is roughly proportional to $\tau^2$ (for
$\tau \simlt 1$), while its position scales as $\ell_{peak}\propto
z_i^{1/2}$ (Zaldarriaga 1997; Fabbri 1999).

To highlight these differences, we plot the normalized discrepancy 
of temperature and polarization angular spectra [defined analogously to 
eq. (1)] between the models in the lower left panel.
Here find that $\Delta C_\ell/C_\ell < 0.07 $ while, around $\ell
\approx 10$, $\Delta P_\ell/P_\ell \approx 1 $.
These results make it clear that although the prompt and simulated
reionization histories share the same value of $z_i$, they produce
noticeably different angular power spectra, and thus correct
constraints on the reionization redshift cannot be obtained by
equating $z_i$ in the prompt and simulated runs.

\subsection{Assigning the Optical Depth}

An alternative approach is to compare our fiducial run with a prompt
model that differs in its $z_i$ but has the same Thomson optical
depth, $\tau=0.08$.  This results in a value for the prompt
reionization redshift of $z_{i,p}=17.1$.  Note that this value is much
larger than the simulated redshift of reionization $z_{i,s}=10.9$, and
corresponds to the redshift at which the simulated ionization fraction
is only $x_{e,s} \approx 0.5$.

\begin{figure}
\centerline{
\epsfig{figure=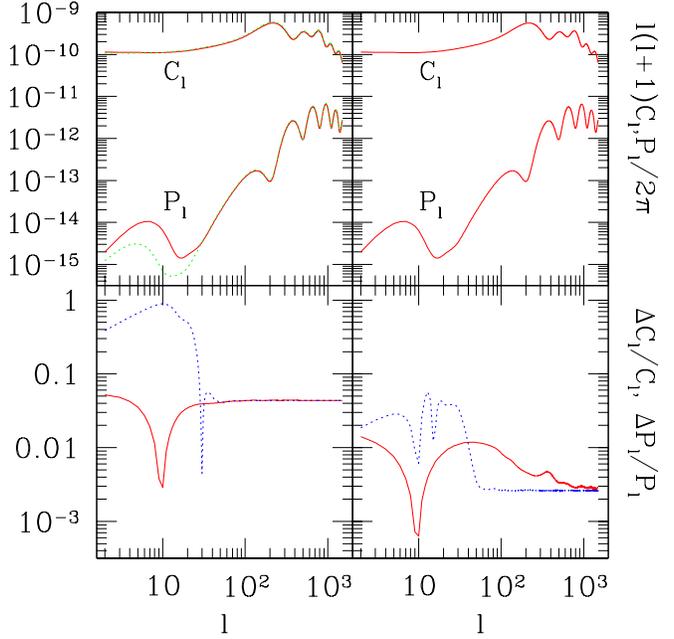, height=10cm}}
\caption{{\protect\footnotesize {{\it Top left}: temperature ($C_\ell$) and 
polarization ($P_\ell$) angular power spectra versus the multipole $\ell$ for 
run A (solid line) and for prompt reionization (dotted line) when 
{\tt CMBFAST} is run with $z_i=10.9$. {\it Bottom left}: the normalized 
temperature and polarization discrepancy as a function of angular multipole 
$\ell$ for this case. Here the solid and dotted lines refer to $C_\ell$ 
and $P_\ell$ respectively.
{\it Top right}: temperature ($C_\ell$) and polarization ($P_\ell$) angular 
power spectra versus the multipole $\ell$ for run A (solid line) and for 
prompt reionization (dotted line) when {\tt CMBFAST} is run with $\tau=0.08$. 
{\it Bottom right}: the normalized temperature and polarization discrepancy 
as a function of angular multipole $\ell$ for this case. Here  the solid 
and dotted lines refer to $C_\ell$ and $P_\ell$ respectively.}}}
\label{fig2}
\end{figure}

Comparing these two runs we find, however, that the discrepancy
between the two visibility functions is only $\Delta g/g \sim
10^{-4}.$ This small change in the visibility function results in an
equally small change in the observed CMB anisotropies.  This can be
seen on the top right panel of in Fig. \ref{fig2} in which the angular
power spectra are almost indistinguishable.  Indeed, the fractional
discrepancy between these runs is only $\Delta C_\ell/C_\ell \lsim
0.01$ and $\Delta P_\ell/P_\ell \lsim 0.06$ as shown in the bottom
right panel in this figure.

Thus the two different reionization models produce comparable angular
power spectra if $\tau$ is specified rather than $z_i$.  In fact if we
try to recover the reionization redshift from the standard formula
$z_i=8.9(\tau/h \Omega_b)^{2/3} \Omega_0^{1/3}$ (Peebles 1993, Tegmark
{\it et al.} 2000) we get $z_i=17.1$,  again a redshift at which $x_e
\approx 0.5$.  It is clear then that by analyzing CMB power spectra
using a prompt model, one can draw reasonable conclusions as to the
overall Thomson optical depth, whereas the reionization redshift is
much more uncertain.

\subsection{Quantification of Errors \& Tests of Our Approach}

\begin{figure}
\centerline{
\epsfig{figure=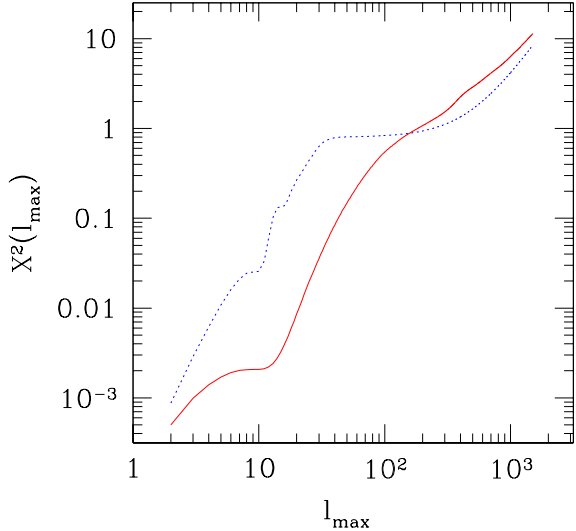, height=7cm}}
\caption{{\protect\footnotesize {$\chi^2$ of $C_\ell$ (solid line) and 
of $P_\ell$ (dotted line) versus $\ell_{\rm max}$ (see eq.\ 2).}}}
\label{fig3}
\end{figure}

In order to quantify the difference between the fiducial model and
the prompt model with the same optical depth further, we plot in Fig.\
\ref{fig3} the quantity
\begin{equation}
\chi^2(\ell_{max}) = \sum_2^{\ell_{max}}\frac{(2\ell+1)}{2}(\frac{\Delta C_\ell}{C_\ell})^{2}
f_{sky}
\label{eq3}
\end{equation}
as a function of the multipole number $\ell_{max}$ assuming full sky
coverage, $f_{sky}= 1$.  For multipole values greater than 100, $\chi^2$
exceeds unity, the value corresponding to the cosmic variance error of
$2C_\ell^2/(2\ell+1)$.  This means that one can find a statistically
significant difference between the prompt and fiducial models only by
looking at angular multipoles with $\ell \gsim 100.$  However, in
Fig. \ref{fig2} we saw that the largest differences between the two
histories occur at lower multipole values ($\ell = 20- 40$).  Thus
although these reionization histories are considerably different, it
is almost impossible to discriminate between them observationally.

Finally, we extend our analysis to the runs B, C, and D of CFGJ,
comparing them with prompt runs with equal values of $\tau_s$.
These results are summarized in Table 1.  As in run A, the
values of $z_{i,p}$ found for these models correspond to 
times at which $x_{e,s} \approx 0.5$, when the IGM was in the midst
of changing from a neutral state to an ionized one.

Is is clear from this table that it is not possible to put a
constraint on the reionization redshift directly from {\tt CMBFAST.}
In fact, as can be easily seen in Table 1, the differences between
$z_{i,s}$ and $z_{i,p}$ cannot be easily parameterized because they
depend on several effects that influence the duration $\Delta z_s$ of
the reionization process.  These effects include radiative and
stellar feedbacks, the star formation efficiency, the photon escape
fraction, the spatial and mass distribution of the ionizing sources,
and even the clumpiness of IGM, which is beyond the scope of the CFGJ
simulations.

We have conducted a number of checks and convergence tests to assess
the robustness of our approach.  Our first check was to examine the
overall, numerical value of the optical depth, which
differs slightly from the input value due to the finite integration 
time step.  Here we found that the output $\tau$s differ by less
than $5 \times 10^{-5}$ from the specified values.  This corresponds
to $\Delta C_\ell/C_\ell \lsim 10^{-4}$, i.e. $30$ times smaller than the
actual difference between the simulated and prompt runs.

Next, we carried out a series of convergence tests, to ascertain the
effects of $k$-mode sampling on our results.  Here we restrict our
tests to run A for which we increased the $k$ resolution by $50\%.$ A
comparison between the original and resampled simulated runs showed
that at most angular scales $\Delta C_\ell/C_\ell$ and $\Delta
P_\ell/P_\ell$ are within $0.1\%$; larger differences (up to $\sim 0.7\%$) are 
found at higher multipole numbers ($\ell \gsim 700$). Comparing
the resampled run A with a resampled prompt run with equal input
optical depth, we found $\Delta C_\ell/C_\ell$ and $\Delta
P_\ell/P_\ell$ values that were virtually indistinguishable from those
obtained with low $k$ resolution runs.

Finally, we examined the effect of time sampling at both $k$
resolutions, comparing two prompt runs with equal optical depth and
$k$ resolution, but forcing one of them to the same time step as in simulated 
run A. Again, we found $\Delta C_\ell/C_\ell$ and $\Delta
P_\ell/P_\ell$ $\lsim 10^{-4}$, far smaller than the discrepancies
between the prompt and simulated runs.

\section{Conclusions}

One of the first stages of nonlinear structure formation, reionization
marked an important transition from a dark and relatively simple
universe to one filled with a dazzling array of stars, galaxies,
quasars, and other nonlinear objects.  And although one of our best
probes of this transition is through the measurement of CMB
fluctuations, the process of reionization itself is much more
dependent on the complicated astrophysical issues important at low
redshifts than the linear issues important at $z\approx 1100.$

In this work, we have explored this transition, quantifying the
impact of realistic scenarios of reionization on the angular power
spectrum of the Comic Microwave Background.  While standard estimates
assume prompt reionization, we have considered instead a range of
simulated models, each with a prolonged reionization epoch.  We find
that equating the redshift of full IGM ionization between these
simulations and models that assume instantaneous reionization leads
to widely discrepant temperature and polarization spectra. On the
other hand, equating prompt and extended models with the same overall
optical depth leads to differences in anisotropies that are nearly
undetectable.  In this case the redshift of complete ionization is
lost in the complicated details of the phase transition, and 
comparisons yield $z_i$ values corresponding to roughly the point of
50\% ionization in the simulations, although even this value is
model dependent.  It is clear then that while $z_i$ is useful
as schematic tool, it is the total optical depth that is most accurate
in providing a definition of the reionization epoch.

\bigskip
MB thanks M. Zaldarriaga for discussions and hospitality
at IAS, Princeton where part of this work has been carried out.
ES has been supported in part by an NSF MPS-DRF fellowship.

\begin{table}
\centerline{Table 1: Reionization parameters}
\begin{center}
\begin{tabular}{|cccccccc|}
\hline \hline
Run&$\Delta z_{s}$$^{(a)}$&$z_{i,s}$$^{(b)}$&$\tau_{s}$$^{(c)}$&
$z_{i,p}$$^{(d)}$ &$x_{e,s}(z_{i,p})$$^{(e)}$&$\frac{\Delta C_\ell}{C_\ell}$&$\frac{\Delta P_\ell}{P_\ell}$\\ \hline \hline
$A$& 15&$ 10.9$&0.080&17.1&0.46&$\lsim 0.01$&$\lsim 0.06$ \\ \hline
$B$& 17&$ 8.3$&0.059&13.9&0.51&$\lsim 0.007$&$\lsim 0.1$ \\ \hline
$C$& 15&$ 15$&0.098&19.8&0.48&$\lsim 0.01$&$\lsim 0.1$ \\ \hline
$D$& 17&$ 8.3$&0.064&14.7&0.47&$\lsim 0.008$&$\lsim 0.1$ \\ \hline \hline
\end{tabular}
\label{tab1}
\caption{{\protect\footnotesize{$^{(a)}$Duration of reionization; 
$^{(b)}$Redshift of 
complete reionization (simulated); $^{(c)}$Thomson Optical Depth (simulated);
$^{(d)}$Redshift of complete reionization (prompt); $^{(e)}$Hydrogen ionization fraction 
at $z_{i,p}$ (simulated).}}} 
\end{center}
\end{table}

\label{lastpage}
\newpage
\end{document}